# Aggregation and magnetism of Cr, Mn, and Fe cations in GaN

N. Gonzalez Szwacki,[1*] J. A. Majewski,[1] and T. Dietl[1,2]

1. *Institute of Theoretical Physics, Faculty of Physics, University of Warsaw, ul. Hoża 69, 00-681 Warszawa, Poland*
2. *Institute of Physics, Polish Academy of Sciences, al. Lotników 32/46, 02-668 Warszawa, Poland*

A first-principles DFT-GGA+$U$ study of the doping of bulk GaN and its surface with Cr, Mn, and Fe confirms a strong tendency of these magnetic ions, occupying Ga sites of the wurtzite and zinc blende phases of GaN, for the formation of embedded clusters. Our study reveals that the tendency for aggregation is larger in the case of Cr and Mn ions than for Fe ones. In contrast to previous DFT-GGA calculations, we predict a ferromagnetic ordering of the Cr and Mn clusters consisting of more than two atoms, whereas it turns out that the Fe clusters are all antiferromagnetic. We have also investigated the magnetic ordering of the nearest-neighbor ionic pairs that substitute gallium atoms at the (0001) wz-GaN surface. We find that Fe ions tend to aggregate, whereas there is a mutual repulsive interaction in the case of Cr and Mn. The nearest neighbor Mn and Fe pairs are coupled antiferromagnetically, whereas the Cr pair is coupled ferromagnetically. Further, the relevance of our findings to recent experimental results is discussed.

# I. INTRODUCTION

Motivated by complementary properties of magnetic and semiconducting systems, a great deal of interest is devoted to demonstrate high temperature ferromagnetism[1,2] or antiferromagnetism[3] in semiconductors. Since the first theoretical prediction of room-temperature ferromagnetism in (Ga,Mn)N containing a sufficiently high concentration of holes in the valence band,[4] GaN-based dilute magnetic semiconductors (DMSs) have been extensively studied experimentally[5-7] and by *ab initio* methods.[8-10]

It was suggested, employing various DFT approaches, that the transition metal (TM) cations in (Ga,TM)N have a strong tendency to aggregate,[11-18] which confirmed the importance of the contribution of highly lying TM *d* orbitals to the cohesive energy. However, the predicted magnetic ground state of particular TM aggregates is a matter of dispute. Furthermore, there was lacking a theoretical study of zinc blende (zb) and wurtzite (wz) GaN doped with Cr, Mn, and Fe that would be performed within the same DFT scheme and using the same computational details. This hindered a reasonable assessment of chemical trends and the role of the crystallographic structure in TM aggregation.

We present here a systematic theoretical investigation of the formation enthalpies and magnetic ordering of small (up to four magnetic ions) Cr, Mn, and Fe aggregates in wz- and zb-GaN employing the DFT-GGA+$U$ framework. In order to gain information on the TM aggregation during the growth, we discuss also magnetic properties of Cr, Mn, and Fe pairs at the (0001) wz-GaN surface. We compare our theoretical results with conclusions of recent experimental works on the magnetic ion distribution and magnetism in (Ga,Cr)N,[19-21] (Ga,Mn)N,[22,23] and (Ga,Fe)N.[24,25]

The paper is organized as follows. In Sec. II the calculation details are presented. In Sec. III we present results for the magnetic clusters of TMs in vacuum. The main body of the present study is contained in Sec. IV, where the results of calculations for TM clusters in bulk GaN and on its surface are discussed. In Sec. V the detailed comparison with existing experimental data is given, whereas the paper is concluded in Sec. VI.

# II. CALCULATIONS DETAILS

The calculations were performed within the DFT-GGA+$U$ approach with the Perdew-Burke-Ernzerhof exchange and correlation functional[26] and ultrasoft pseudopotentials[27] to describe electron-ion interactions as employed in the plane wave basis Quantum Espresso[28] code. The kinetic energy cut off was set to 30 Ry (180 Ry for the electronic density) for the calculation involving TM in GaN and it was increased by a factor of two for the calculations of isolated TM dimers. A possibility of non-collinear magnetism has not been taken into account in our calculations that does not include spin-orbit interactions. It worth noting, however, that the presence of a cycloidal spin structure has been predicted for (Ga,Mn)As thin layers in the presence of spin-orbit interactions.[29]

For GGA+$U$ calculations, we adopted the values of the $U$ parameter equal to 3.0, 3.9, and 4.3 eV for Cr, Mn, and Fe 3*d* orbitals, respectively. Similar values of $U$ had been used to describe the properties of $CrO_2$,[30] (Ga,Mn)As,[31] (Ga,Mn)N,[13,31] FeO,[32] and (Ga,Fe)N.[33] We assumed that the Stoner exchange parameter, $J$, is set to zero, or alternatively that its effects can be mimicked redefining the $U$ parameter as $U_{eff} = U - J$. For details on the implementation of the GGA+$U$ approach in the Quantum-Espresso package we refer to Ref. 34.

We used the supercell geometry for calculations of TMs in the bulk GaN, isolated TMs dimers, and TMs at the GaN surface. To investigate the formation of TM clusters, we used supercells consisting of 64 ($2a \times 2a \times 2a$) and 96 ($2\sqrt{3}\,a \times 3a \times 2c$) atoms for the zb and wz lattices, respectively, and correspondingly the suitable $4 \times 4 \times 4$ and $2 \times 2 \times 2$ Monkhorst-Pack $k$-point sampling meshes were employed. In the case of TM dimers, to avoid interactions between the periodically repeated dimers, the calculations were carried out in a large cubic unit cell of side 15 Å, and we adopted a single $k$-point ($\Gamma$-point) sampling in the Brillouin zone (BZ) integration. Finally, the surface modeling was carried out using a supercell containing four GaN bilayers (48 Ga and 48 N atoms) and a vacuum region of approximately 10 Å. The first GaN bilayer was fixed in the appropriate bulk optimize configuration. The (0001) GaN surface unit cell has been taken to be $2\sqrt{3}\,a \times 3a$, and the dangling bonds on the opposite nitrogen surface were saturated with twelve H atoms to help reduce finite fields that could be produced across the supercell. The integration over the BZ was performed using the $4 \times 4 \times 1$ Monkhorst-Pack $k$-point mesh.

The atomic positions in the supercells were fully optimized in all cases. Within the employed computational scheme, the lattice constant of zb-GaN was calculated to be 4.549 Å, whereas for wz-GaN the lattice constants $a = 3.211$ Å and $c = 5.231$ Å were obtained, which agree fairly well with other theoretical calculations[10, 13] and experimental data.[35]

To give a broader perspective on to the computational methods used to describe the properties of solids with TMs we should mention that in recent papers[10, 13, 36] a different strategy was used than the one adopted in this paper to deal with the self-interaction error. Mixing an exact (Hartree Fock) exchange with the DFT exchange-correlation functional gives a hybrid DFT/HF approach which, although computationally very expensive, is in many aspects more accurate than DFT alone. To have a taste on the performance of hybrid functionals in their ability to predict the properties of extended systems, we refer the reader to Ref. 37.

Finally, we note that in recent studies[10, 38] the presence of a Jahn-Teller distortion around Mn in a cubic and wurtzite GaN was predicted. Since we optimize atomic positions, such a distortion is probably also present in our calculations but it has not been systematically investigated in this study.

### III. ISOLATED MAGNETIC CLUSTERS

The performed studies of the TMs in GaN were based on the DFT-GGA+$U$ scheme. To check the difference between outcomes of the DFT-GGA and the DFT-GGA+$U$ methods, we decided to make a calculation for isolated TM dimers, which are known to be rather challenging objects for *ab initio* methods.[39] Therefore, we do not expect to have a quantitative agreement with experimental results, but rather to gain information about possible inaccuracies and differences between theoretical results for $U = 0$ and $U \neq 0$ cases.

For each TM dimer, we calculated the dissociation energy, bond length, magnetic moment, and the HOMO-LUMO energy gap employing GGA and GGA+$U$ schemes. The results are summarized and compared to available experimental data in Table I. As can be seen, the GGA+$U$ results reproduce correctly the magnetic properties of the dimers, since the calculated magnetic moments are in perfect agreement with the experimental data. For Mn$_2$, the GGA+$U$ scheme, comparing to the GGA method, clearly improves the values for the bond length and dissociation energy. For Fe$_2$, both GGA with and without $U$ tend to overestimate the dissociation energy but give satisfactory results for the bond length.

Finally, the GGA results for Cr$_2$ are in disagreement with the experimental ones and the $U$ correction does not improve the situation, the fact already noted in the previous theoretical studies.[40] We also calculated the vibration frequencies within the GGA+$U$ method, which are equal to 194 (481), 81 (76.4), and 340 (300) cm$^{-1}$ for Cr$_2$, Mn$_2$, and Fe$_2$, respectively, with experimental values taken from Ref. 41 and depicted in parentheses. Again (like in the case of the bond length), a rather large disagreement is observed for the Cr$_2$ dimer.

A similar result for Fe$_2$ has been obtained by Rollmann et al..[42] The authors have shown that explicit consideration of the on-site Coulomb repulsion within the GGA+$U$ has a strong effect on the potential energy surface of Fe$_2$ and also larger Fe clusters.

It should be noted that the properties of small isolated TM$_n$, $n > 2$, clusters were also studied both experimentally and theoretically.[43-45] For instance, the Stern–Gerlach experiments on Mn$_n$ clusters with $n = 5$ to 22 revealed a ferrimagnetic nature of the clusters whose magnetic moment oscillates from 0.4 to 1.7$\mu_B$/atom.[44] The density functional theory calculations were able to reproduce these data. For example, Bobadova-Parvanova et al.[45] predicted that for small Mn$_n$ ($n < 6$) clusters, the antiferromagnetic (AF) and ferromagnetic (FM) ordering of atomic magnetic moments would be energetically nearly equivalent, whereas the AF arrangement would be clearly favored for larger clusters. This type of magnetic behavior could be, however, altered by the presence of nitrogen atoms. Indeed, Rao et al.[46] studied the magnetic moments of nitrogen-doped Mn clusters and found that the coupling between the magnetic moments at Mn sites would remain ferromagnetic irrespective of their size or shape.

## IV.   MAGNETIC CLUSTERS IN THE GaN MATRIX

### A.   The studied structures

We investigated the energetics and magnetism of the TM ions in the zinc blende and wurtzite GaN matrices by occupying up to four Ga cation sites of the suitable supercells with TM ions and looked for their energetically most favorable positions. The substitutional position of the TM ions in the GaN matrix is well documented experimentally by EXAFS studies.[22, 33] Since the formation of Cr and Mn magnetic clusters in the wz-GaN matrix seemed to be also well documented in the literature, we took as the starting point the energetically most stable structures found in Refs. 12 and 11 for Cr and Mn, respectively. For two Cr and Mn atoms, the most stable configuration is when these atoms occupy the nearest-neighbor (NN) Ga sites in the (0001) plane. For three Cr and Mn atoms in the supercell, the most favorable configuration is the one where all three atoms occupy nearest-neighbor Ga sites in the (0001) plane. Finally, for four atoms in the supercell, the lowest energy configuration is when the TM atoms form a pyramid like structure with three TM atoms in the (0001) plane and one out of plane. All these geometrical arrangements of TM atoms are depicted in Fig. 1. We are not aware about theoretical studies devoted to the formation of Fe$_n$ clusters in wz-GaN. However, as it was mentioned above, the isolated iron clusters were extensively studied[43] and the energetically most favored structures for small clusters were found to be similar to those mentioned above for Cr and Mn in wz-GaN. Therefore, we assumed that also the Fe$_3$ and Fe$_4$ clusters would have a triangle and pyramid like structure, respectively, in the wz-GaN matrix.

For comparison purposes, the same magnetic clusters have been studied in the zb-GaN matrix (see Fig. 1). Note that in zinc blende crystallographic phase the <111> directions are equivalent to the [0001] direction in wurtzite lattice. A similar theoretical study was done for TM clustering in zb-GaN[14] and more recently in GaAs.[47]

The bond lengths between TM in the considered $TM_n$ clusters in GaN (see Fig. 1) and the energetically most favorable spin configurations are listed in Table II. For comparison, we have also included the Ga-Ga distances in undoped wz- and zb-GaN.

### B. The tendency for clustering

Let us start the discussion of the results by considering the tendency for the formation of Cr, Mn, and Fe clusters in GaN. We begin by considering the heat of the reaction,

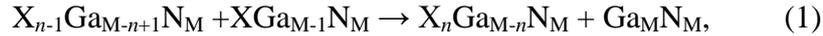

$$X_{n-1}Ga_{M-n+1}N_M + XGa_{M-1}N_M \rightarrow X_nGa_{M-n}N_M + Ga_MN_M, \quad (1)$$

where $M = 32$ or $48$ for the zb and wz GaN lattices, respectively, and X denotes Cr, Mn, or Fe. For $n = 2$ this heat of reaction is usually called the TM-TM pair interaction energy or pairing energy.[47] In Fig. 2 we plot the heat of reaction defined in Eq. (1) as a function of the number of TM atoms in the cluster. The plotted energies clearly show that Cr, Mn, or Fe ions prefer to be close to each other occupying neighboring Ga sites, which confirms the previous theoretical studies.[11-18]

Two important conclusions can be drawn from the findings presented in Fig. 2. First, the tendency for the formation of clusters is larger for the Cr and Mn atoms than for the Fe atoms. Second, in the case of $Fe_n$ clusters, the values for the heat of reaction are very close for both host matrixes.

In order to demonstrate the critical role of magnetic effects in the magnitude of the pairing energy we have also carried out the calculations with no spin polarization, as shown in Fig. 3. A qualitative difference between results presented in Figs. 2 and 3 points to the importance of both chemical and magnetic energies in determining the tendency towards clustering. When the spin polarization is switched off the tendency for clustering essentially increases with the number of TM-TM bonds involving $d$ orbitals formed in the $TM_n$ cluster upon addition of one TM atom to the $TM_{n-1}$ cluster. This is especially clear for the zb-GaN case in Fig. 3. Generally, the presence of the spin polarization in the system leads to a more complex picture with a lower tendency for clustering and smaller variations with the cluster size.

### C. Magnetic ordering

In Fig. 4 we show the total energy difference between FM and AFM ordering of the magnetic moments for the Cr, Mn, and Fe dopants in GaN versus the number of atoms in the magnetic cluster. The FM spin order is favorable for Cr and Mn impurities, whereas the AFM ordering is expected in the case of Fe doping. We can see in Fig. 4 that the results for Mn and Fe are not sensitive to the crystal structure of the matrix (wz vs. zb), whereas for Cr the picture is somewhat different, i.e., our results predict FM ordering for the wz-GaN host and the transition from FM to AFM ordering for the $Cr_n$ clusters in the zb-GaN case. It should be pointed out that in contrast, for instance, to (Ga,TM)As (Ref. 47) the magnetic interaction between the TM ions is short ranged in GaN and limited virtually only to NNs.[48]

The dependence of the magnetic moments per atom is shown in Fig. 5 for $Cr_n$, $Mn_n$, and $Fe_n$ clusters. For $n = 1$ in all cases the self-consistent magnetic moments assume integer values of 3, 4, and $5\mu_B$ for Cr, Mn, and Fe, respectively. For larger Mn clusters, the value of the magnetic moment per ion saturates and remains equal to $4\mu_B$. Similarly, for Cr clusters in wz-GaN, the net magnetic moment per ion is always equal to $3\mu_B$. In Fig. 5, we can also see that the magnetic moment of the Fe clusters depends on the cluster size, but not on the crystallographic structure of the GaN matrix. Interestingly, for the $Cr_4$ clusters, the magnetic moment differs in the wz- and zb-GaN, just exhibiting its dependence on the number of Cr atoms in the cluster in the zb-GaN matrix.

### D. TM clustering at the (0001) wz-GaN surface

We have also calculated the pairing energy of the TM atoms on the Ga (0001) surface of wz-GaN. Interestingly, Fe cations have a negative pairing energy, equal to -125 meV, close in value to that in bulk (see Fig. 2). This means that during epitaxial growth the Fe atoms have a tendency to aggregate at the growing Ga surface. The Cr and Mn pairs, instead, have positive pairing energies of 284 and 173 meV, respectively, which points to a more homogenous distribution of these TM atoms during the epitaxial growth.

The magnetic moments of isolated TM atoms are 6, 5 and $4\mu_B$ for Cr, Mn, and Fe, respectively. The net magnetic moment of a system with one Ga replaced by a TM atom at the (0001) wz-GaN surface is 4.55, 4.44, and $3.86\mu_B$ for Cr, Mn, and Fe, respectively. This means that the value of the net magnetic moment induced by Fe at the surface is closer to the value corresponding to a free atom than to the net magnetic moment in bulk GaN (see Fig. 5 for $n = 1$). The net magnetic moments of Cr and Mn atoms at the wz-GaN surface are in between the values corresponding to the free atoms and the ion in bulk GaN.

The magnetic interaction of TM pairs at the Ga surface is different from that shown in Fig. 4 for the bulk. The Fe pair is antiferromagnetically coupled both at the surface as in the bulk, whereas the magnetic coupling of two Mn (Cr) ions that substitute nearest neighbor Ga atoms is AFM (FM) at the surface and FM (FM) in bulk. Similar results have been reported for Mn (Cr) pairs in bulk and at a mixed gallium-nitrogen ($11\bar{2}0$) wz-GaN surface[49] (for Cr see Ref. 50), and also for Mn-incorporated GaN(0001) surface.[51]

### V. COMPARISON TO EXPERIMENTAL RESULTS

The determined here and in previous *ab initio* studies magnetic moments clearly indicate that the Ga-substitutional TM impurities assume a high spin 3+ charge state in GaN, in agreement with the experimental studies for dilute and uncompensated by residual donors (Ga,Cr)N,[19, 21] (Ga,Mn)N,[22, 52, 53] and (Ga,Fe)N[54, 55] systems.

In accord with the *ab initio* studies, the presence of a strong FM coupling between single Ga-substitutional NN Mn pairs was experimentally demonstrated for (Ga,Mn)N.[23] There are indications that FM interactions dominate also for larger Mn clusters,[23] substantiating the results displayed in Fig. 4. The underlying mechanism is presumably a ferromagnetic super-exchange, examined theoretically a time ago for the case of NN pairs in Cr-doped II-VI compounds[56] and more recently for NN Mn pairs in GaAs.[57]

A robust high temperature ferromagnetism was discovered in epitaxial wz-(Ga,Cr)N.[19, 21] Transmission electron microscopy and x-ray diffraction studies of these samples reveal the presence of Cr-rich nanocrystals formed mostly by chemical phase separation, as the content of secondary crystallographic phases is small.[58] A FM character of Cr-rich wz-(Ga,Cr)N nanocrystals corroborates our findings. We note that while bulk CrN is antiferromagnetic, epitaxial strain can turn it into a high temperature ferromagnet.[59]

The chemical phase separation was also detected in wz-(Ga,Fe)N.[54, 60] However, in the Fe case, Fe-rich wz-(Ga,Fe)N nanocrystals co-exist with precipitates of other compounds such as ε-$Fe_3N$. According to subsequent detail studies,[24, 25] $Fe_xN$ precipitates with $x > 2$ are ferromagnetic. However, when the degree of nitridation is enlarged by increasing the growth temperature,[24] so that $x \leq 2$, the dominant interactions become antiferromagnetic, as predicted here for Ga-substitutional $Fe_n$ clusters in wz-GaN ($x = 1$).

It is worth noting that the accumulated experimental results point to the formation of $Fe_xN$ nanocrystals at the growth surface and their segregation towards the surface during the epitaxy,[24, 25, 60] provided that the growth rate is sufficiently small.[60] These observations are in accord with our findings which point to a large gain in energy by aggregation of Fe cations at the surface (Sec. IV). At the same time, the annealing up to 900°C does not affect Fe distribution.[24] This is consistent with our results, despite that they point to a similar pairing energies at the surface and in the bulk, as the kinetic barriers for diffusion are typically several orders of magnitude larger in the bulk case.

In contrast to Fe, no aggregation of Mn cations is expected at the surface according to our data (Sec. IV). Indeed, comparing to Fe,[24] about one order of magnitude greater concentrations of diluted Mn cations can be incorporated into GaN.[23] At the same time, as shown in Fig. 2, a sizable negative pairing energy for NN Mn cations in bulk GaN is predicted within our theory. Nevertheless, (Ga,Mn)N synchrotron x-ray diffraction spectra were found insensitive of annealing up to 900°C.[22] This points to the presence of a relatively large magnitude of a kinetic barrier for Mn diffusion in GaN.

Particularly intriguing is the case of (Ga,Cr)N. Within our approach Cr cations are not expected to aggregate at the surface (Sec. IV) but a strongly negative pairing energy is expected for Cr pairs in bulk GaN. On the experimental side, the frequently observed high temperature ferromagnetism[19] points to an important role of Cr aggregation. This may mean that our model underestimates the tendency towards pairing or that kinetic barriers are relatively low for Cr diffusion in GaN.

## VI. CONCLUSIONS

In summary, we have investigated the coupling of magnetic Cr, Mn, and Fe ions substituting Ga sites in wz and zb-GaN matrices by first principles DFT-GGA+$U$ calculations and we have confirmed a strong tendency of these magnetic ions for clustering in GaN. The magnetic ordering of the Fe clusters is always AFM in contrast to that of the Mn and Cr clusters which is FM (with the exception of $Cr_4$ in zb-GaN that is AFM). We have also found a tendency for clustering of Fe pairs at the (0001) wz-GaN surface, where the magnetic interaction between Fe ions is, like in the bulk, still AFM. We predict that for the Mn pairs the magnetic interaction at the surface is AFM, i.e., it has changed from FM in bulk to AFM at the surface. Our studies allow for successful interpretation of available experimental data and shed light on physical mechanisms governing the observed phenomena. We have also underlined the need of information on TM diffusion for a fuller description and deeper understanding of the TM aggregation in semiconductors.

**ACKNOWLEDGMENTS**

The work was supported by the European Research Council through the FunDMS Advanced Grant within the "Ideas" Seventh Framework Programme of the EC and InTechFun (POIG.01.03.01-00-159/08) of EC. We thank Alberta Bonanni and Maciej Sawicki for valuable discussions.

TABLE I. Comparison between theoretical (GGA and GGA+$U$) and experimental values of dissociation energy, bond length, magnetic moment, and HOMO-LUMO gap for isolated $Cr_2$, $Mn_2$, and $Fe_2$ dimers.

|  | GGA | GGA+$U$ | exp. |
|---|---|---|---|
| $Cr_2$ ||||
| dissociation energy | 1.06 | 0.60 | 1.56[a] |
| bond length | 2.227 | 2.706 | 1.679[b] |
| magnetic moment | 0 | 0 | 0[b] |
| HOMO-LUMO gap | 1.69 | 2.44 | - |
| $Mn_2$ ||||
| dissociation energy | 0.69 | 0.19 | 0.1-0.56[c] |
| bond length | 2.592 | 3.318 | 3.4[b] |
| magnetic moment | 10 | 0 | 0[b] |
| HOMO-LUMO gap | 0.57 | 2.09 | - |
| $Fe_2$ ||||
| dissociation energy | 2.55 | 2.35 | 1.15[d] |
| bond length | 2.035 | 2.135 | 2.02[b] |
| magnetic moment | 6 | 8 | 8[e] |
| HOMO-LUMO gap | 0.03 | 0.90 | - |

[a]Ref. 40, [b]Ref. 41, [c]Ref. 46, [d]Ref. 61, [e]Ref. 39

TABLE II. The TM-TM distances for the TM$_n$ clusters inside the wz- and zb-GaN matrices. The numbers in curly brackets refer to the positions of TM ions, as shown in Fig. 1. The magnetic ordering is specified with arrows, and the sequence of arrows corresponds to the numbering from Fig. 1. The nearest neighbor Ga-Ga distances for the gallium sublattice in wz- and zb-GaN are also included. All values of bond lengths are in angstroms.

|    | $n = 2$ | $n = 3$ | $n = 4$ |
|----|---------|---------|---------|
|    | GaN | | |
| wz | {1-2}=3.21 | {1-2}=3.21 | {1-2}=3.21 |
| zb | {1-2}=3.22 | {1-2}=3.22 | {1-2}=3.22 |
|    | (Ga,Cr)N | | |
| wz | ↑↑<br>{1-2}=3.07 | ↑↑↑<br>{1-2}=3.12 | ↑↑↑↑<br>{1-2}=3.07, {1-4}=3.09 |
| zb | ↑↑<br>{1-2}=3.08 | ↑↑↑<br>{1-2}=3.10 | ↑↑↑↓<br>{1-2, 1-4}=2.93 |
|    | (Ga,Mn)N | | |
| wz | ↑↑<br>{1-2}=3.16 | ↑↑↑<br>{1-2}=3.17 | ↑↑↑↑<br>{1-2}=3.17, {1-4}=3.25 |
| zb | ↑↑<br>{1-2}=3.19 | ↑↑↑<br>{1-2}=3.22 | ↑↑↑↑<br>{1-2}=3.28 |
|    | (Ga,Fe)N | | |
| wz | ↑↓<br>{1-2}=3.17 | ↑↑↓<br>{1-2}=3.21<br>{1-3, 2-3}=3.17 | ↑↓↓↑<br>{1-2, 1-3, 2-4, 3-4}=3.16<br>{1-4}=3.21, {2-3}=3.19 |
| zb | ↑↓<br>{1-2}=3.18 | ↑↑↓<br>{1-2}=3.22<br>{1-3, 2-3}=3.17 | ↑↓↓↑<br>{1-2, 1-3, 2-4, 3-4}=3.17<br>{1-4, 2-3}=3.21 |

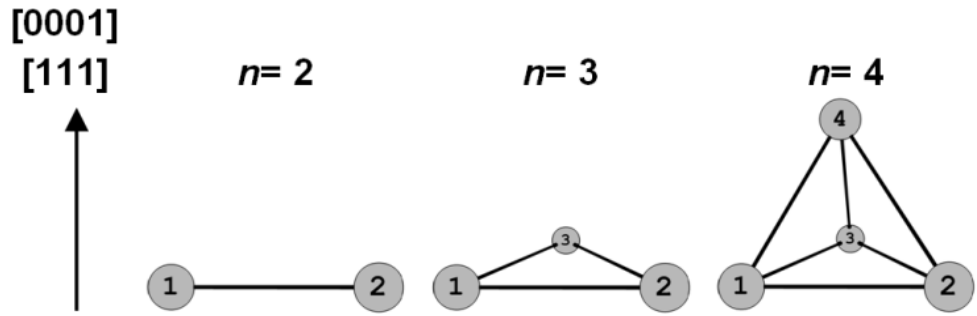

Fig. 1 (Color on line) Geometries of the TM$_n$ ($n$ = 2,3, and 4) clusters in wurtzite and zinc blende GaN, which were studied in the paper. The gray circles represent TM substituting the Ga cations.

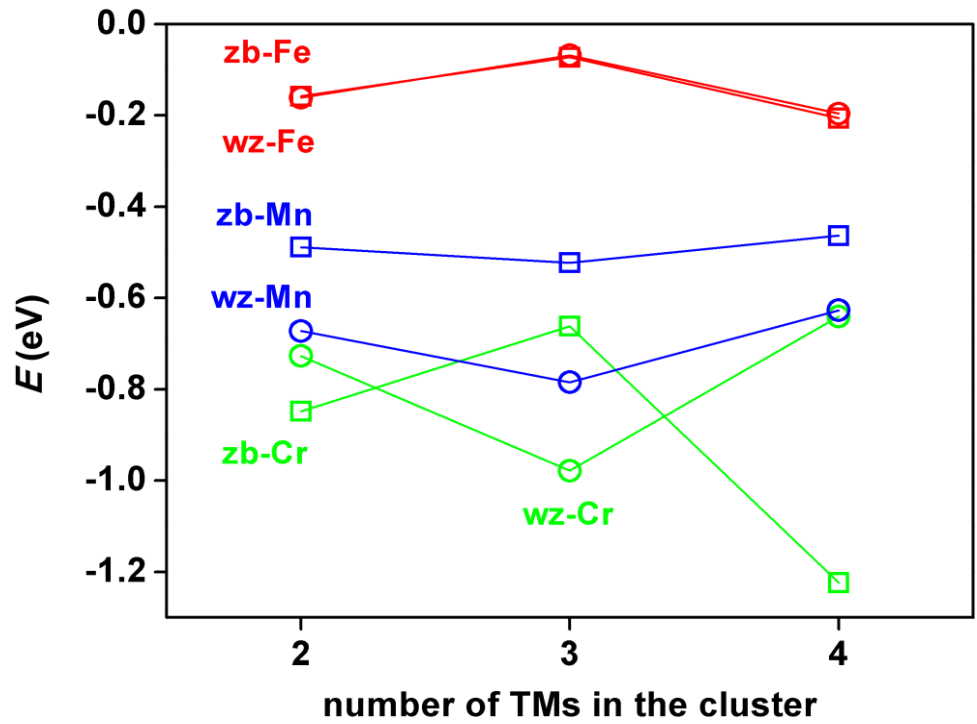

Fig. 2 (Color on line) Heat of reaction as a function of the number of TM atoms in the Cr, Mn, and Fe clusters in the wz- and zb-GaN.

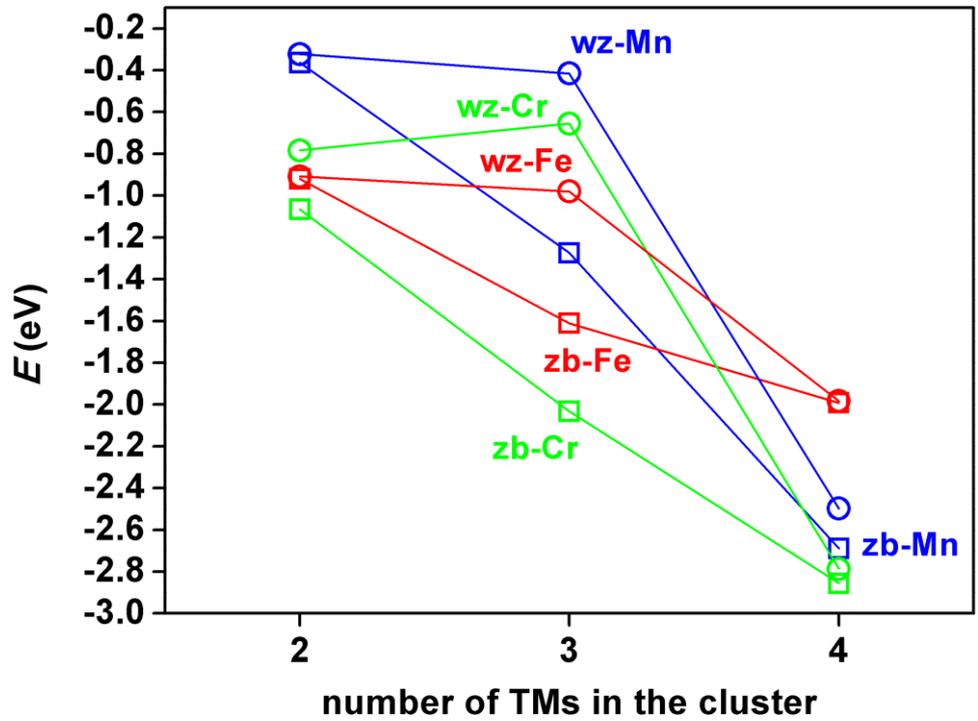

Fig. 3 (Color on line) Heat of reaction as a function of the number of TM atoms in the Cr, Mn, and Fe clusters in the wz- and zb-GaN, calculated without taking into account the spin polarization. This exhibits pure chemical aspect of the clustering.

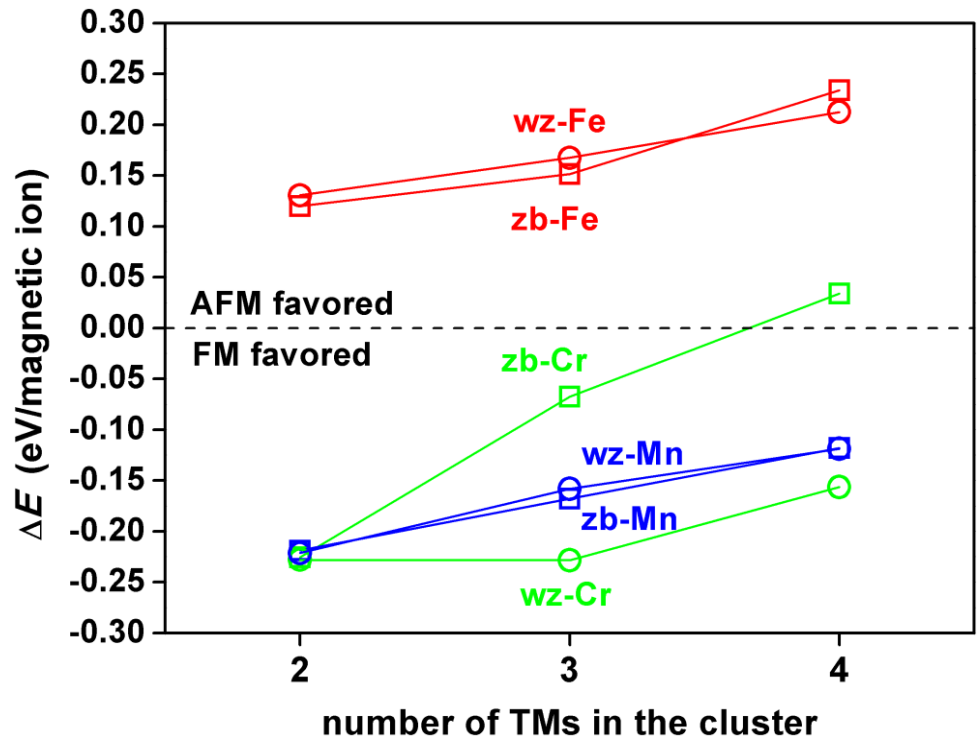

Fig. 4 (Color on line) Energy difference per TM atom between the FM and AFM configurations as a function of the number of TM atoms in the Cr, Mn, and Fe clusters in the wz- and zb-GaN.

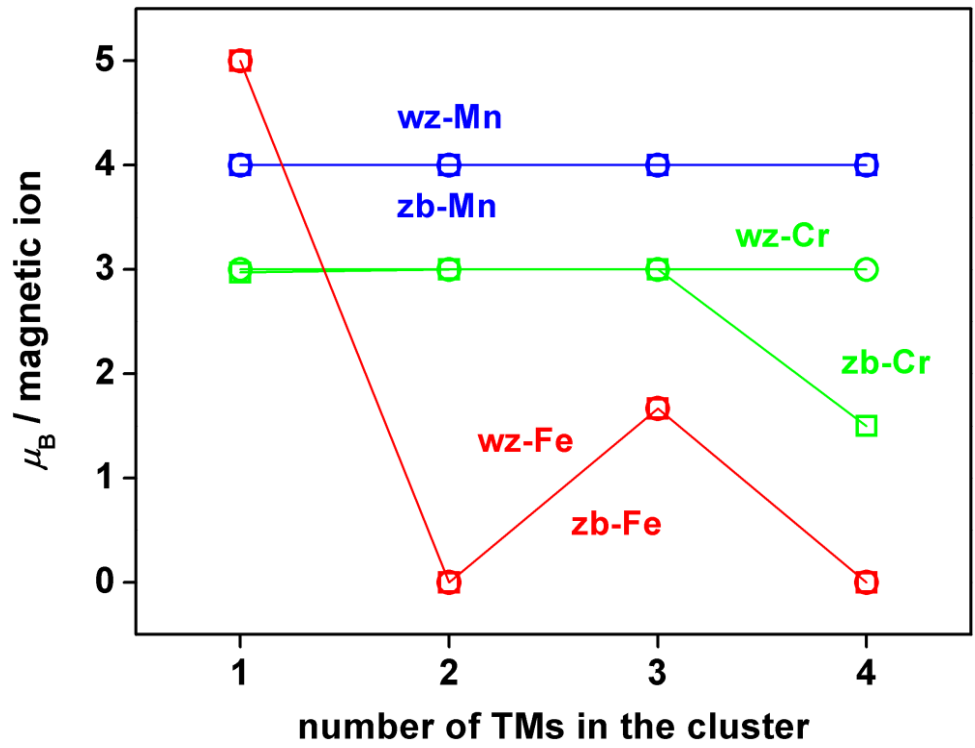

Fig. 5 (Color on line) Net magnetic moment (in Bohr magnetons) as a function of the number of TM atoms in the Cr, Mn, and Fe clusters in the wz- and zb-GaN.